\documentstyle[twoside,fleqn,espcrc2]{article}

\newcommand {\beq}{\begin{equation}}
\newcommand {\eeq}{\end{equation}}
\newcommand {\beqa}{\begin{eqnarray}}
\newcommand {\eeqa}{\end{eqnarray}}
\newcommand {\beqan}{\begin{eqnarray*}}
\newcommand {\eeqan}{\end{eqnarray*}}

\newcommand {\Romannumeral}[1]{\uppercase\expandafter{\romannumeral#1}}

\newcommand {\ee}{\mbox{e}}

\newcommand{\AmS}{{\protect\the\textfont2
  A\kern-.1667em\lower.5ex\hbox{M}\kern-.125emS}}

\title{Applications of the overlap formalism to super Yang-Mills
theories}

\author{J. Nishimura\address{Department of Physics, Nagoya
University,\\
Furo-cho, Chikusa-ku, Nagoya 464-01, Japan}}%
       
\begin{document}

\begin{abstract}
We show that the idea to use the overlap formalism to 
formulate 4D ${\cal N}=1$ super Yang-Mills theory on the lattice
without fine-tuning can be applied to 3D ${\cal N}=1$ case as well.
Another application we propose is 
a regularization of the IIB matrix model,
which is large $N$ reduced model of 10D ${\cal N}=1$ super
Yang-Mills theory.
\end{abstract}

\maketitle

\section{Introduction}
Recent analytical developments in supersymmetric gauge theories
\cite{SW,SUSY} motivate an approach from the lattice formulation to 
these theories.
Supersymmetry on the lattice, however, is naturally a hard problem
since the lattice breaks the continuous rotational and translational 
invariance, and the latter is a subgroup of the supersymmetry. 
However, as is the case with other symmetries such as chiral symmetry
in QCD, one can always recover the symmetry in the continuum limit
by fine-tuning.
This is the reconciliation proposed by Curci and Veneziano \cite{CV},
who showed within lattice perturbation theory that
4D ${\cal N}=1$ super Yang-Mills theory can be obtained
by using the Wilson-Majorana fermion for the gluino
and fine-tuning the hopping parameter to the chiral limit.
Based on this work, some numerical simulations have been started
\cite{num}.

In general, thanks to the universality of field theory,
one can hope to obtain a supersymmetric theory by fine-tuning
as many parameters as the relevant operators that break supersymmetry
around the supersymmetric ultraviolet fixed point.
If we have some symmetry that forbids those supersymmetry breaking operators,
we can avoid fine-tuning by imposing the symmetry on the lattice theory.
This is analogous to how we deal with the continuous rotational and 
translational invariance,
which can be restored only in the continuum limit, 
but without any fine-tuning, so long as we maintain the discrete rotational 
and translational invariance on the lattice.

The overlap formalism \cite{NN}, which has been originally developed to
deal with chiral gauge theories, is useful for the above purpose
since it preserves the symmetry of massless fermions
manifestly on the lattice.
In Ref. \cite{NN}, it has been suggested that the overlap formalism
can be used to formulate
4D ${\cal N}=1$ super Yang-Mills theory on the lattice without fine-tuning.
A method using the domain-wall formalism, which is more suitable
for numerical simulation, has been proposed in Ref. \cite{4DN1}

In this article, we review the application of 
the overlap formalism to 3D ${\cal N}=1$ super Yang-Mills theory
\cite{superYM}
as well as to a regularization \cite{unitary} of the IIB matrix model, 
which is nothing but the large $N$ reduced
model of 10D ${\cal N}=1$ super Yang-Mills theory.

\section{3D ${\cal N}=1$ super Yang-Mills theory}
When we consider super Yang-Mills theory in three dimensions,
we should note first that in odd dimensions, 
the gluon can acquire mass without violating
gauge invariance through Chern-Simons term.
Supersymmetry then only requires the gluon and the gluino mass to be equal.
Here we consider only the massless case,
which can be obtained by imposing the parity invariance 
since it prohibits both the gluon and the gluino mass.
In Wilson fermion formalism, the Wilson term breaks the parity
invariance and one has to fine-tune the hopping parameter in order to 
make the fermion massless.
The whole theory is still not parity invariant, however,
since the Chern-Simons term is generated through
the fermion loop \cite{CL}. This is known as parity anomaly.
One can add a Chern-Simons term as a counterterm to obtain a parity
invariant theory, though this is only possible when
the Chern-Simons term to be added has a proper coefficient
which ensures the invariance of the theory under large gauge
transformations.
For SU($N$) gauge group, the present case with Majorana
fermion in the adjoint representation requires
that the $N$ should be even
so that we may impose both parity invariance and gauge invariance.

In Ref. \cite{parity}, we have pointed out that formulating
Dirac fermion in odd dimensions in a parity invariant way
is exactly the same problem as that of formulating chiral gauge theory in
even dimensions. The overlap formalism, therefore, must be useful,
and indeed, we have shown that the formalism gives 
a lattice regularization of massless Dirac fermion
in which the fermion determinant is parity invariant and real.
The formalism as well as the above statements
has been generalized to Majorana fermion \cite{superYM}.
The overlap being real but not necessarily positive leaves the place
for the global gauge anomaly, which has been found to reproduce the 
correct result expected in the continuum \cite{KN}.
In the present case with Majorana fermion in the adjoint
representation, the $N$ of the gauge group SU($N$) should be even
so that the theory may be free from the global gauge anomaly.
The exact parity invariance ensures that the continuum limit is
supersymmetric.
Thus the overlap formalism makes 3D super Yang-Mills theory
accessible without fine-tuning and without dealing with the
Chern-Simon term on the lattice, in contrast to the Wilson fermion
approach.

\section{Application to IIB matrix model}
Another application of the overlap formalism we propose \cite{unitary}
is a regularization of the IIB matrix model \cite{IKKT},
which is proposed as a nonperturbative formulation
of type IIB superstring theory.
The eigenvalues of the 10 $N \times N$ hermitian matrices $A_\mu$
are interpreted as 10D space-time coordinates and
the U(1)$^{10}$ symmetry:
$A_\mu \rightarrow A_\mu + \alpha_\mu {\bf 1}$ corresponds to 
the 10D translational invariance of the space time.
However, the regularization adopted by Ref. \cite{IKKT}:
$| \mbox{eigenvalues of }A_\mu | \le \frac{\pi}{a}$
violates the U(1)$^{10}$ symmetry.
A natural way to regularize the theory
without violating this symmetry is to replace the hermitian
matrices by unitary matrices: $U_\mu = \ee^{ia A_\mu}$.
It is now the phases of the eigenvalues that are 
interpreted as the space-time coordinates and
the space time is naturally compactified to a ten-dimensional torus.
However, since now we are
essentially dealing with a large $N$ reduced model of a lattice
theory, we will have problems with the doublers and the supersymmetry.
Since the fermion is chiral, the elimination of the doublers is
non-trivial.
Here we use the overlap formalism.
We note that although the overlap formalism as a regularization of 
ordinary lattice chiral gauge theories has a subtle problem with
the local gauge invariance not being preserved on the lattice,
its application to the present case is completely safe regarding this,
since we do not have the local gauge invariance to take care of.
The global gauge invariance, on the other hand, 
which is indeed one of the important symmetry of the model,
is manifestly preserved within the formalism.
The unitary matrix model thus defined has the U(1)$^{10}$ symmetry:
$U_\mu \rightarrow \ee^{i \theta _\mu} U_\mu$,
which corresponds to the 10D translational invariance of the 
space time.

Another important symmetry of the IKKT model is the ${\cal N}$=2 
supersymmetry regarding the eigenvalues of the bosonic Hermitian matrices 
as the space-time coordinates \cite{IKKT}.
Note that the 10D translational invariance mentioned above
forms a subgroup of the supersymmetry.
This ${\cal N}$=2 supersymmetry comes from (1) the supersymmetry 
of 10D super Yang-Mills theory, combined with (2) 
the symmetry under constant shifts
of the fermionic matrices.
When we consider the unitary matrix model with the overlap formalism,
we have (2) but not (1) unfortunately, and therefore,
the ${\cal N}$=2 supersymmetry is not manifest.
However, we adopt a similar spirit as in the case with ordinary 
super Yang-Mills theories
and expect the ${\cal N}$=2 supersymmetry to be restored in the 
double scaling limit without particular fine-tuning.

We can calculate 
one-loop effective action around BPS-saturated states analytically.
For example, the classical vacuum is given by $N$ D-instanton
configuration, which is BPS-saturated.
We find that the logarithmic attractive potential
between two D-instantons
induced by the integration over the 
bosonic degrees of freedom is cancelled
by the contribution from the fermionic degrees of freedom when
the D-instantons are sufficiently close to each other.
On the other hand, when they are farther apart, a weak attractive
potential arises, unlike the case with IKKT model, where the potential is
completely flat.
Thus the U(1)$^{10}$ symmetry of the unitary matrix model
is spontaneously broken in the weak coupling limit, 
though much more mildly than 
in purely bosonic case, where the logarithmic attractive potential 
exists.

Since the U(1)$^{10}$ symmetry is expected to be restored
in the strong coupling phase, there must be a phase transition.
We consider that this phase transition provides a natural place 
to take the double scaling limit.
Since our model preserves manifest U(1)$^{10}$ symmetry,
we can discuss the dynamical generation of the space time
as the spontaneous breakdown of the U(1)$^{10}$ symmetry.
Roughly speaking, if the U(1)$^{10}$ symmetry is broken down to
U(1)$^{D}$ for sufficiently small string coupling constant, 
our model is equivalent to a $D$-dimensional gauge theory
due to the argument of Eguchi-Kawai \cite{EK}.
However, according to Ref. \cite{FKKT}, we have to take the double
scaling limit in the weak coupling region.
Since gauge theories in more than four dimensions
should have an ultraviolet fixed point in the strong coupling regime, if any,
the U(1)$^{10}$ symmetry of our model is expected to be broken down 
at least to U(1)$^{4}$ for sufficiently small string coupling constant.
This gives a qualitative understanding of the dynamical origin of
the space-time dimension 4.

\section{Future Prospects}
It is interesting to examine if the idea to apply the overlap formalism
to super Yang-Mills theories works for the series of
super Yang-Mills theories \cite{BSS}
obtained through dimensional reduction of
6D ${\cal N}=1$ and 10D ${\cal N}=1$ theories.
Here, the parent theories suffer from gauge anomaly,
but the theories after dimensional reduction are anomaly free.
Four-dimensional ${\cal N}=2$ super Yang-Mills theory \cite{SW}
might be accessible without fine-tuning in this way. 
Also of interest is to see whether there exists a
nontrivial fixed point in more than four dimensions
for supersymmetric case, since the conclusion for purely bosonic
case turned out to be negative \cite{KNO,largeN}.
This issue is worth addressing also in the context of nonperturbative
formulation of superstring theory using matrix models \cite{unitary,IKKT}.
Numerical simulations of IIB matrix models must be extremely
exciting, since we may be able to understand why the space-time dimension
is four.


\begin{thebibliography}{9}
\bibitem{SW} 
N. Seiberg and E. Witten, Nucl. Phys. B426 (1994) 19; B431 (1994) 484. 
\bibitem{SUSY} N. Seiberg, Nucl. Phys. B435 (1995) 129; K. Intriligator and 
N. Seiberg, Nucl. Phys. Proc. Suppl. 45BC (1996) 1.
\bibitem{CV} G. Curci and G. Veneziano, Nucl. Phys. B292 (1987) 555.
\bibitem{num} See I. Montvay's article in this volume.
\bibitem{NN} R. Narayanan and H. Neuberger, Nucl. Phys. B443 (1995) 305.
\bibitem{4DN1} J. Nishimura, Phys. Lett. B406 (1997) 215;
See also T. Hotta's article in this volume.
\bibitem{superYM} 
N. Maru and J. Nishimura, preprint, hep-th/9705152.
\bibitem{unitary}
N. Kitsunezaki and J. Nishimura, preprint, hep-th/9707162.
\bibitem{CL} A. Coste and M. L\"uscher, Nucl. Phys. B323 (1989) 631.
\bibitem{parity} R. Narayanan and J. Nishimura, to be published in
Nucl. Phys. B, hep-th/9703109.
\bibitem{KN} Y. Kikukawa and H. Neuberger, preprint, hep-lat/9707016.
\bibitem{IKKT} 
N. Ishibashi, H. Kawai, Y. Kitazawa and A. Tsuchiya,
Nucl. Phys. B498 (1997) 467.
\bibitem{EK} 
T. Eguchi and H. Kawai,
Phys. Rev. Lett. {\bf 48} (1982) 1063.
\bibitem{FKKT} 
M. Fukuma, H. Kawai, Y. Kitazawa and A. Tsuchiya,
preprint, hep-th/9705128.
\bibitem{BSS} L. Brink, J.H. Schwarz and J. Scherk, 
Nucl. Phys. B121 (1977) 77.
\bibitem{KNO} H. Kawai, M. Nio, and Y. Okamoto,
Prog. Theor. Phys. 88, (1992) 341.
\bibitem{largeN} J. Nishimura, Mod. Phys. Lett. A11 (1996) 3049.
\end{thebibliography}
\end{document}